\documentclass[useAMS,usenatbib]{mn2e}
\usepackage{epsfig}

\usepackage{enumerate}
\usepackage{journal_names}

\def\Re{\mbox{$R_{\rm e}$}}
\def\Msun{\mbox{$M_\odot$}}

\def\Ysun{\mbox{$\Upsilon_\odot$}}
\def\ML{\mbox{$M/L$}}
\def\LCDM{\mbox{$\Lambda$CDM}}
\def\kms{\mbox{km~s$^{-1}$}}

\def\Ystar{\mbox{$\Upsilon_{\star}$}}

\def\vir{\mbox{$_{\rm vir}$}}

\def\cvir{\mbox{$c_{\rm vir}$}}

\def\Ysol{\mbox{$\Upsilon_{\odot, V}$}} 

\def\lsim{\mathrel{\rlap{\lower3.5pt\hbox{\hskip0.5pt$\sim$}}
    \raise0.5pt\hbox{$<$}}}                
\def\gsim{~\rlap{$>$}{\lower 1.0ex\hbox{$\sim$}}}

\title[Mass, anisotropy and IMF of NGC 5846]{The SLUGGS Survey: Breaking degeneracies between dark matter, anisotropy and the IMF using globular cluster subpopulations in the giant elliptical NGC 5846}

\author[Napolitano et al.]{\noindent
Nicola~R.~Napolitano$^{1}$\thanks{E-mail: napolita@na.astro.it}, Vincenzo~Pota$^{2}$, Aaron~J.~Romanowsky$^{3}$, \and Duncan~A.~Forbes$^{2}$, Jean~P.~Brodie$^{4}$ and Caroline Foster$^{5}$
\\~\\
$^1$ INAF -- Osservatorio Astronomico di Capodimonte, Salita Moiariello, 16, 80131 - Napoli, Italy\\
$^2$ Centre for Astrophysics \& Supercomputing, Swinburne University, Hawthorn VIC 3122, Australia\\
$^3$ Department of Physics and Astronomy, San Jos\'e State University, One Washington Square, San Jose, CA 95192, USA\\
$^4$ University of California Observatories, 1156 High Street, Santa Cruz, CA 95064, USA\\
$^5$ Australian Astronomical Observatory, PO Box 915, North Ryde, NSW 1670, Australia\\}

\begin{document}
\date{Accepted  Received }
\pagerange{\pageref{firstpage}--\pageref{lastpage}} \pubyear{xxxx}
\maketitle

\begin{abstract}
We study the mass and anisotropy distribution of the giant elliptical galaxy NGC~5846 using stars, as well as the red and blue globular cluster (GC) subpopulations. We break degeneracies in the dynamical models by taking advantage of the different phase space distributions of the two GC subpopulations to unambiguously constrain the mass of the galaxy and the anisotropy of the GC system. Red GCs show the same spatial distribution and behaviour as the starlight, whereas blue GCs have a shallower density profile, a larger velocity dispersion and a lower kurtosis, all of which suggest a different orbital distribution. We use a dispersion--kurtosis Jeans analysis and find that the solutions of separate analyses for the two GC subpopulations overlap in the halo parameter space. The solution converges on a massive dark matter halo, consistent with  expectations from \LCDM\ and WMAP7 cosmology in terms of virial mass ($\log M_{\rm DM} \sim13.3\Msun$) and  concentration ($\cvir\sim8$). This is the first such analysis that solves the dynamics of the different GC subpopulations in a self-consistent manner. Our method improves the uncertainties on the halo parameter determination by a factor of two and opens new avenues for the use of elliptical galaxy dynamics as tests of predictions from cosmological simulations.  
The implied stellar mass-to-light ratio derived from the dynamical modelling is fully consistent with a Salpeter initial mass function (IMF) and rules out a bottom light IMF. The different GC subpopulations show markedly distinct orbital distributions at large radii, with red GCs having an anisotropy parameter $\beta\sim0.4$ outside $\sim3\Re$ ($\Re$ is the effective radius), and the blue GCs having $\beta\sim0.15$ at the same radii, while centrally ($\sim1\Re$) they are both isotropic. We discuss the implications of our findings within the two--phase formation scenario for early-type galaxies. 
\end{abstract}


\section{Introduction}\label{sec:intro}

Early-type galaxies are key laboratories where the dark matter (DM) paradigm can be tested. They are among the 
most massive stellar systems in the universe for which accurate kinematical measurements are available. These include integrated long slit stellar kinematics \citep[e.g.,][]{Kronawitter,Gerhard01,Jardel11,Salinas12}, integral field 3D kinematics \citep{Cappellari13} and discrete kinematic tracers such as planetary nebulae (PNe, \citealt{Romanowsky03}; \citealt{Napolitano}, N+09 hereafter; \citealt{Napolitano11}, N+11) and globular clusters \citep[GCs, e.g., ][]{Cote03,Romanowsky09,Schuberth,Richtler11,Norris12}. The latter probe the galaxy projected kinematics (\citealt{2001A&A...377..784N}) out to large galactocentric distances \citep{2004ApJ...602..685P,2004ApJ...602..705P}. These radii are comparable to those explored in the late 70s in spiral galaxies, in studies which resulted in the initial discovery of DM \citep{Rubin,Bosma}. X-rays are often used as viable mass probes, although the number of systems with the necessary undisturbed gas distributions is limited (e.g., \citealt{2006ApJ...646..899H,2012ApJ...755..166H}). 

A growing body of evidence shows that many different stellar subsystems within galaxies share the same phase space properties, i.e. starlight, red GCs (RGCs) and PNe (see e.g. \citealt{Coccato} and \citealt{Pota13}, P+13 hereafter). Other tracers (e.g. the blue GCs, BGCs) are clearly decoupled, despite (presumably) sharing the same potential. Indeed, it is thought that BGCs may trace the galaxy halo component more closely than these other tracers \citep{2006ARA&A..44..193B,Forbes12}.

In general, these studies find good agreement between the RGCs and the underlying starlight,  a connection that is corroborated by the consistency with other discrete stellar tracers like PNe (P+13), although notable exceptions exist \citep{Foster11}. If BGCs are instead governed by the overall galaxy potential in a manner distinct from the RGCs, they act as an independent tracer of the potential to large galactocentric distances (up to $\ge$10 effective radii, \Re). This provides strong constraints on the dark matter distribution with radius. Thus, the combination of RGCs and BGCs offers the opportunity to break the well-known degeneracy between mass and orbital anisotropy that has so far plagued dynamical studies of early-type galaxies (\citealt{1990AJ.....99.1548M}).

Earlier studies have claimed that hot stellar systems are extremely difficult to model by means of discrete dynamical tracers \citep{Merritt91}. However, it has now become clear that, in the case of quasi-spherical systems, the
combination of velocity maps from discrete tracers (e.g. GCs and/or PNe) and the kinematical information from starlight actually provides a powerful tool for constraining the galaxy mass distribution \citep{Napolitano02,Romanowsky09} as well as the orbital distribution (\citealt{Saglia00}; N+09; \citealt{2009MNRAS.395...76D}).

Notwithstanding their intrinsic robustness, these studies have yet to solve two basic problems: 1) identifying the initial mass function (IMF), and 2) breaking the DM halo parameter degeneracies.

The IMF is the prime source of uncertainty in determining stellar masses as it can translate into variations in the stellar mass-to-light ratio ($M/L$) of a factor of two or more. Evidence for systematic variations of the IMF among galaxies has been obtained by measuring the direct effect of the giant-to-dwarf star ratio on integrated galaxy spectra via spectral indices (e.g. \citealt{2012ApJ...760...71C,Spiniello+12,2013MNRAS.429L..15F}) and from studies that combine galaxy dynamics and lensing to infer the ``gravitational'' stellar mass-to-light ratio, (\Ystar\, e.g., \citealt{Thomas+11,2012Natur.484..485C,2013MNRAS.432.2496D,Tortora+13}). In order to quantify the IMF variations, these spectroscopic and dynamical studies have measured the ratio of \Ystar\ to that of the Milky Way, $\Upsilon_{\rm MW}$ ($\delta_{\rm IMF}=\Ystar/\Upsilon_{\rm MW}$), with the result that more massive galaxies were found to have systematically larger $\delta_{\rm IMF}$ (see \citealt{Tortora+13} for a review). However, the above dynamical studies relied solely on data from the inner regions of galaxies. Here, all assumed a fixed DM halo density distribution \citep[e.g., from collisionless simulations,][NFW]{NFW} and the total mass (e.g. from abundance matching studies, \citealt{Moster+10}) and the only free parameters involved was the stellar $M/L$. As the DM quantities need to be tested against observations (e.g., more radially extended dynamical studies such as the one we present here), the argument in the dynamical studies carried out to date is thus somewhat circular. By contrast, kinematic tracers such as PNe or GCs, which extend out to large galactocentric radii, will allow a direct and self-consistent assessment of both the IMF and the DM halo in galaxies (see N+11).
     
The other problem is that the intrinsic/physical relationship between the various DM halo parameters, e.g. the concentration--virial mass \citep{2006ApJ...646..899H} or the halo density--scale radius (N+11) is well aligned with the typical confidence contours of the modeled parameters. This makes the overall analysis highly degenerate.
For this reason, a combined analysis of the different tracers (i.e., stars, RGCs and BGCs in our case) in a single galaxy may help to break the degeneracies in the typical halo parameter space. Modeled values for the different tracers may have confidence contours that are tilted with respect to each other in halo parameter space and converge to a narrower common parameter volume. This would significantly reduce the locus of allowed halo solutions. It would also make early-type galaxies (ETGs) a powerful cosmological testbed for separating different DM species (e.g. Cold from Warm) as they are expected to occupy different areas of the parameter space \citep[see e.g.,][]{Schneider12}. Simultaneous modelling of different GC subpopulations has been attempted for a few cases and has so far yielded contradictory results \citep{Schuberth,Schuberth12}. The implication may be that the assumption that both GC subpopulations are in dynamical equilibrium is incorrect \citep{Schuberth}. Alternatively, due to the inherent difficulties associated with cleanly separating RGCs from BGCs (P+13), the BGC kinematics may have been poorly constrained.


Here we analyze stars plus RGC and BGC subpopulations of the giant elliptical galaxy NGC~5846 and perform a self-consistent dynamical analysis in order to simultaneously and self-consistently constrain the stellar $M/L$, the halo parameters in the context of \LCDM\ predictions, and the orbital anisotropy of the galaxy. With the method outlined above, we attempt to resolve typical degeneracies, in particular that between halo mass and anisotropy.

NGC~5846 is the central and brightest galaxy in a group \citep{Mahdavi05} 23.1 Mpc away (\citealt{Tonry01}\footnote{Distance modulus corrected by -0.16  as per \citet{Jensen03}.}). It is nearly spherical in shape with Hubble type E0 and is kinematically classified as a slow rotator \citep{Emsellem11}. These properties indicate that it is suitable for mass modelling under the assumption of spherical symmetry.

The GC system was studied using Hubble Space Telescope images \citep{Forbes97B, Chies-Santos06}. As in most massive galaxies, the GC system is clearly bimodal in colour with BGC (metal-poor) and RGC (metal-rich) subpopulations as for most massive galaxies. Combining the HST with Subaru images, P+13 showed that the RGCs are more radially concentrated than the BGCs. Using Keck spectroscopy of GCs, they measured little or no rotation for the RGCs (in agreement with the starlight), and systematically higher velocity dispersion in the BGCs than the RGCs. 

Mass profile modelling of NGC~5846 has been performed previously in \citet[][K+00 hereafter]{Kronawitter} and \citet[][2013b]{Cappellari07} who used the kinematics of the galaxy starlight to probe the mass within $\sim$1 effective radius. In order to probe larger galactocentric distances, \citet{Saxton10} combined starlight kinematics with PNe, \citet{Das} used X-rays and PNe, and \citet{DeasonB} used PNe. 

The paper is organised as follows: In \S\ref{sec:data} we present the spatial distribution and kinematics of the GC system for both red and blue GC subpopulations. The dispersion--kurtosis procedure is presented in \S\ref{sec:dynamics}, while its results are derived in \S\ref{sec:models} for the RGCs and BGCs, separately in the first instance, and then jointly in order to find a single equilibrium potential that can account for all observed kinematics. A discussion of the results and our conclusions can be found in \S\ref{sec:conclus}.

\section{Data}\label{sec:data}
We analyse the GC dataset presented in P+13. These data were acquired as part of the SAGES Legacy Unifying Globulars and GalaxiesS (SLUGGS) survey. The survey combines wide-field imaging with Keck/DEIMOS \citep{2003SPIE.4841.1657F} multi-object spectroscopy of globular clusters and galaxy starlight for a sample of 25 nearby early-type galaxies\footnote{Further details of the survey are available in Brodie et al. (2014, in prep.) and at http://sluggs.ucolick.org}.

For NGC~5846, $HST/WFPC2$ \citep{Forbes97B,Chies-Santos06} and new Subaru $gri$ photometry is available for the photometric pre-selection of GC candidates (see P+13 for details). The spectroscopic sample of P+13 consists of 195 confirmed GCs collected over six DEIMOS masks. As mentioned above, the GC system exhibits the classical colour bimodality with a colour split between BGCs and RGCs around $(g-i)=0.95$. For the time being, the separation is based on the photometry only, although we will also consider this separation from a dynamical point of view.

\begin{figure}
\hspace{-0.6cm}
\psfig{file=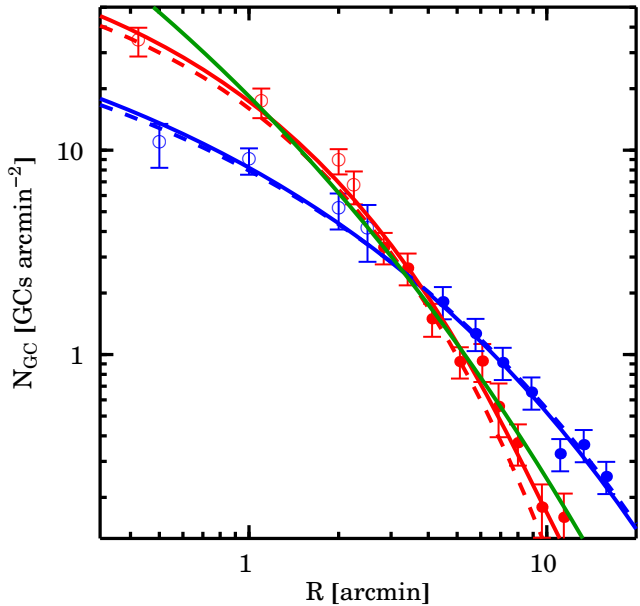,width=0.5\textwidth}
\caption{Sersic fit to the stellar surface brightness (SB, green line) as compared to the RGC (red) and BGC (blue) surface density profiles. The open and filled circles correspond to HST and Subaru data, respectively. The surface brightness has been arbitrarily scaled to match the red $N_{\rm GC}$. It is evident that the RGCs closely follow the stellar SB within the uncertainties, although the fit to the RGCs (red solid line) deviates slightly at large radii (see discussion in the text). The Sersic fit to the BGCs number density profile is shown as a blue solid line. Using a different colour cut to separate RGCs from BGCs (see \S\ref{sec:kinem}) does not significantly affect the slope of the density profiles as shown by the red and blue dashed lines (also arbitrarily scaled to match the normalisation of the respective solid lines).}\label{fig:fig_SB}
\end{figure}

\subsection{Density profiles}\label{sec:density}
Fig. \ref{fig:fig_SB} shows the radial number density profiles ($N_{\rm GC}/{\rm arcmin}^2$) of the photometrically selected RGCs and BGCs together with the $V-$band galaxy surface brightness (SB) profile from K+00 adopted as the reference starlight radial distribution. We show the  \citet{1963BAAA....6...41S} fit to the stellar SB profile ($n=4.9$, $\Re=230''$ and  $\mu_0=14.3$ mag/arcsec$^{-2}$) extrapolated over the whole radial range. The fit is vertically offset to match the RGC number density data in the outskirts. As a result, the match within $R\sim1'$ is not optimal. The RGC number density profile is depleted with respect to the stellar SB profile in the inner regions due to GC cluster disruption. 
We assume $\Re=81''$ as per \citet{Cappellari06} and \citet{DeasonB} henceforth, which corresponds to 9.1 kpc at the distance of NGC~5846. The RGCs closely follow the normalised stellar light distribution outside of $1'$. A S{\'e}rsic fit to the RGC number density distribution yields $n=2.9$, $\Re=160''$ and $N_e=3.3$ GC arcsec$^{-2}$.

The similarity between the radial RGC number density and the stellar SB profiles indicates a close connection between the RGCs and the overall stellar population as also discussed in e.g. \citet{Forbes2768} and P+13. Remaining small deviations at large radii might be attributable to uncertainties on the GC background determination (see Eq. 1 of P+13, here $bg=$0.13 arcmin$^{-2}$). These minimal offsets are insignificant compared to the model uncertainties. This motivates the use of the stellar SB profile to derive the de-projected stellar density profile for input into the Jeans equations, similar to the method employed for other discrete dynamical tracers (e.g. PNe, N+09; N+11).


Since NGC~5846 is a non-rotating E1 galaxy (\citealt{Coccato}), we expect the deviation from spherical symmetry to be less than 10\% \citep{Kronawitter, Binney}. Under the assumption of spherical symmetry, the deprojection of the observed SB is unique and can be derived as usual via the Abel integration \citep{Binney}:
\begin{equation}\label{eq:abel}
j(r)=\frac{1}{\pi}\int_0^r\frac{dI}{dr}\frac{dr}{\sqrt{r^2+R^2}},
\end{equation}
where $r$ and $R$ are the 3D and 2D radii, respectively, and $I(R)$ is the interpolated SB in units of $L_{\sun}/$arcsec$^2$. Eq. \ref{eq:abel} is solved numerically. The same procedure is adopted to derive the 3D distribution of the BGCs, this time by fitting the S{\'e}rsic profile to the BGC 2D distribution in Fig. \ref{fig:fig_SB} and then applying Eq. \ref{eq:abel} to derive the 3D density. 

We derive the total stellar luminosity by integrating the extrapolated SB profile and obtain $L_{tot}=9.8\times10^{10}L_{\odot,V}$. We then derive the total stellar mass by assuming a 
constant stellar $M/L$ (\Ystar in solar units), via the simple equation $M_{\rm star}=\Ystar \times L_{tot}$. The stellar density is given by $\rho_\star(r)=\Ystar \times j(r)$. In particular, we fit \Ystar\ as a free parameter in the dynamical analysis and compare it with the same quantity derived from Stellar Population Synthesis (SPS) models (see \S\ref{sec:ML}) in order to compute the IMF mismatch parameter ($\delta_{\rm IMF}$). 

The S{\'e}rsic fit to the 2D density of the BGCs shown in Fig. \ref{fig:fig_SB} has an overall more diffuse profile than the one of RGCs, i.e. $N_e=0.24$ GC arcsec$^{-2}$, $n=2.9$ and $\Re=780''$. We have checked that the number density profile of the BGCs is not sensitive to the colour cut adopted in the photometricly selected sample in P+13. As we shall see in \S\ref{sec:kinem}, the kinematics of the two subpopulations suggest that one can refine the colour selection of the two subpopulations, but this does not significantly impact the number density slope, as shown by the dashed lines in Fig. \ref{fig:fig_SB}, where we have taken RGC from $g-i>1.0$ and BGC from $g-i<0.85$. Thus, we keep the BGC number density profile obtained using a colour separation at $g-i=0.95$ in order to minimise the uncertainties.

\subsection{Mass-to-light ratio from SPS models}\label{sec:ML}

One key parameter we want to address in the dynamical analysis is the stellar $M/L$ as
derived from dynamical models. This will allow us to evaluate the most likely IMF compatible with the best fit dark matter halo and the observed kinematics. Here we derive the SPS based $M/L$ based on literature stellar parameters.

According to the analysis of absorption line indices presented in \citet{Denicolo}, stars in NGC~5846 are old (11.7 Gyr) and metal-rich ([Fe/H] = 0.19). Using this information, we 
estimate a plausible stellar $\ML$ using the \citet{Worthey94} models (i.e. the same as adopted by \citealt{Denicolo}) and obtain $\ML_{\rm V}=7.2$ assuming a \citet{Salpeter55} IMF and solar $(M/L_{V})_\odot=4.83$.

This central $M/L$ estimate is slightly lower than that adopted by SAURON in the $I-$band. Their estimate was $\ML_{\rm I}=3.33$ based on the SPS analysis within $\sim1 R_e$ (\citealt{Cappellari06}, using a \citealt{Kroupa01} IMF). For $V-I=1.28$\footnote{from Hyperleda: http://leda.univ-lyon1.fr/.}, this corresponds to $\ML_{\rm V}=8.7$ for a \citet{Salpeter55} IMF or $\ML_{\rm V}=5.4$ for a Kroupa IMF.

The $M/L$ has recently been revised to $\ML_{\rm R}=7.13$  in ATLAS$^{\rm 3D}$ 
\citep{2013MNRAS.432.1862C} using the \citet{2012MNRAS.424..157V} models and a Salpeter IMF, which also corresponds to $\ML_{\rm V}=8.6$ (Salpeter IMF) or $\ML_{\rm V}=5.9$ (Kroupa IMF). 

As our reference value, we have chosen the average of all $V$-band estimates given above, i.e. 8.1$\pm0.9$\Ysun (Salpeter IMF) or 5.1$\pm0.6$\Ysun (Kroupa IMF) and 4.6$\pm0.5$\Ysun (Chabrier IMF). 

\begin{figure}
\vspace{-0.4cm}
\hspace{-0.7cm}
\psfig{file=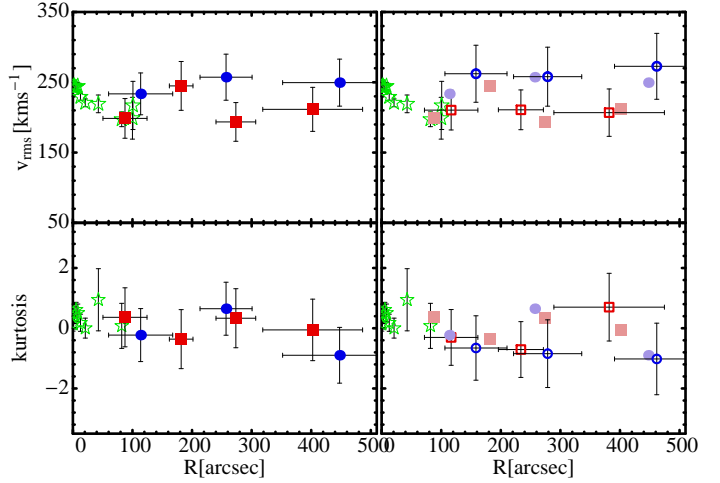,width=0.53\textwidth}
\caption{Dispersion and kurtosis profiles for the RGC (red symbols), BGC (blue symbols) and folded longslit stellar kinematics from K+00 (green stars). The left and right panels show the analysis for a single colour separation at $g-i=0.95$  and the differential colour cut presented in \S\ref{sec:kinem}, respectively. For ease of comparison, light blue and red filled symbols in the right panel replicate the original $g-i=0.95$ cut separation from the left panel.}\label{fig:fig_kin}
\end{figure}

\subsection{Kinematics}\label{sec:kinem}

In Fig. \ref{fig:fig_kin} we show the velocity dispersion profile of the RGCs and BGCs separated according to P+13 using a colour threshold $g-i=0.95$, together with the long slit data from K+00 after averaging the positive and negative data at a corresponding radius. We also show the excess kurtosis profile (simply kurtosis henceforth) of the two subpopulations, which is a measure of the deviation from Gaussianity of the velocity distribution, and hence a proxy for orbital anisotropy. For the stellar measurement, we convert the $h_4$ parameter from K+00 according to the usual conversion formula $\kappa\simeq 8\sqrt{6} h_4$ \citep{1993ApJ...407..525V}. For GCs we use an unbiased kurtosis estimator (\citealt{J&G95}, also see the Appendix for a discussion) tested on PN samples (e.g., N+09). The GC velocity and velocity dispersion profiles shown above are slightly different from the ones reported in P+13 because we used a different binning in order to include more GCs in each spatial bin to derive reliable kurtosis estimates. The red and blue GC samples used for the kurtosis estimate are smaller than samples used in other analyses (e.g. N+09 and N+11), although we ensure that each radial bin contains at least 30 GCs. However, this requirement is slightly relaxed for the more stringent colour selections (see below) where radial bins contain $\gsim$20 GCs. We have  checked that there are no biases in the kurtosis estimates due to small number statistics using a suite of Monte Carlo simulations (see the Appendix). We find that the adopted definition reliably recovers the intrinsic kurtosis of simulated systems despite large statistical uncertainties.

As previously stressed in P+13, there is good agreement between the measured velocity dispersion and kurtosis of the innermost RGC and the stellar data. This confirms the working assumption that RGCs are tracers of the underlying stellar population, which we adopt hereafter.

In contrast, RGCs and BGCs have markedly different dispersion profiles with the outermost data points differing by $\sim 30~\kms$ despite agreeing in the central regions. As for the kurtosis profiles, both GC subpopulations agree at all radii. However and unlike the density profiles (see \S\ref{sec:density}),  the kinematic measurements are more sensitive to the reciprocal contamination between sub-populations as evidenced in figure 5 of P+13, wherein it is clear that the two subpopulations are too mixed around the chosen colour threshold. Thus, we explore the effects of an improved colour separation of the red and blue GC in order to reduce the probability of cross-contamination. In particular, we investigate whether cross-contamination depends on radius as a result of relative fractions (see Fig. \ref{fig:fig_SB}) of RGCs and BGCs that dominate inside and outside $R\sim250''$, respectively. 

\begin{figure}
\hspace{-0.4cm}
\psfig{file=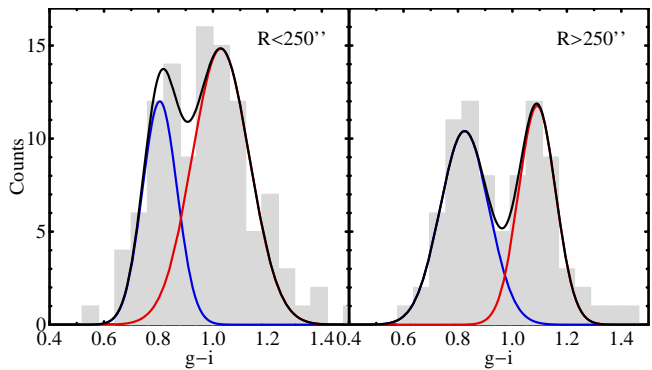,width=0.55\textwidth}
\caption{
Colour separation of the RGCs and BGCs for inner (left) and outer (right) radial bins as labelled. The two subpopulations are separated using a double gaussian fit (black line). Individual components are shown in red and blue for the RGCs and BGCs, respectively. The relative fractions and mean colours of the two subpopulations are different between the two radial bins as expected from their differing number density distributions (Fig. \ref{fig:fig_SB}). 
}\label{fig:fig_histo}
\end{figure}

In Fig. \ref{fig:fig_histo} we show the GC colour distribution for two radial bins. The bimodal GC colour distribution is fitted with a double gaussian. We find that the relative fractions of RGCs and BGCs vary with radii (also see e.g., \citealt{1998MNRAS.293..325F}). In the inner radial bin (i.e. $R<250''$), the optimal colour separation for minimal contamination of the RGC sample is $g-i>1.0$, while the purest sample of BGCs corresponds to $g-i<0.8$. Clean samples in the outer bin are selected with $g-i>1.05$ and $g-i<0.85$ for the RGCs and BGCs, respectively.

The drawback of adopting this exclusive selection approach is an obvious reduction in sample size with which to compute the kinematics. Indeed, this differential colour selection yields a final sample of $\sim140$ GCs, a significant decrease from the original 195. This approach is necessary for reliably delineating the behaviour of the two subpopulations in phase space as demonstrated in Fig. \ref{fig:fig_kin}, where we compute the kinematics using differential colour cuts. In this figure, the differences in the kinematics of the two subpopulations are enhanced thanks to the differential colour cut. The dispersion profile of the BGCs is systematically larger than that of the RGCs by $\sim 40~\kms$. The kurtosis profiles of the two subpopulations diverge beyond $R\sim250''$ with RGCs reaching up to $\sim1$ and the BGCs plummeting to as low as $\sim -1$. These differences are significant at the $\sim2\sigma$ level.

Whether this results from differences in the respective anisotropy distributions of the two subpopulations is not straightforward and will be investigated as part of the dynamical analysis presented in \S\ref{sec:dynamics}. We first present dynamical arguments based on a simple algebraical approach and evaluate the connection between the observed kinematics and possible orbital distribution of the two GC subpopulations in the next section.

\subsection{Effect of tracer density profiles on the observed kinematics}\label{sec:effect}

As a prelude to the dynamical analysis to be presented in \S\ref{sec:dynamics}, we quantify the possible impact of the distinct number density profiles (see  \S\ref{sec:density}) on the measured kinematics of the RGC and BGC. 
The radial component of the velocity dispersion is given by the radial Jeans equation:
\begin{equation}
v_{\rm circ}^2(r) = \sigma_r^2(r) \,[\alpha(r)+\gamma(r)-2\beta(r)], 
\label{eq:jeansrad}
\end{equation}
where $v_{\rm circ}^2(r)=GM(r)/r$, $\alpha\equiv -d\ln j/d\ln r$, $\gamma\equiv -d\ln \sigma^2/d\ln r$ and
$\beta=1-\sigma_\theta^2/\sigma_r^2$, with $\sigma_\theta$ and $\sigma_r$ the azimuthal and radial dispersion components in spherical coordinates, respectively. Assuming a power-law profile $v_{\rm circ}^2=V_0^2r^{-\gamma}$ $\sigma^2=\sigma_0^2 r^{-\gamma}$, $\alpha$, $\gamma$ and $\beta$ are constant with $r$ and the projected dispersion can be written as \citep{Dekel05}:
\begin{equation}
\sigma_p^2(R)= A(\alpha,\gamma)\, 
\left(\frac{(\alpha+\gamma)-(\alpha+\gamma-1)\beta}
{(\alpha+\gamma)-2\beta}\right)
V_0^2\,  R ^{-\gamma},
\label{eq:sig_p}
\end{equation}
where
\begin{equation}
A(\alpha,\gamma)= \frac{1}{(\alpha+\gamma)}
\frac{\Gamma[(\alpha+\gamma-1)/2]}{\Gamma[(\alpha+\gamma)/2]}
\frac{\Gamma[ \alpha/2]}{\Gamma[(\alpha-1)/2]}  \ .
\end{equation}

While we are clearly not in the simple power-law regime, the slope of the number density profiles mimics a power-law profile with changes of $\pm10\%$ within the radial range covered by the kinematic data. Thus the use of Eq. \ref{eq:sig_p} is a reasonable approximation for the arguments below (see also \citealt{Dekel05}).
The SB fit presented in \S\ref{sec:density} corresponds to $\alpha_{\rm RGC}\sim3.1$ and $\alpha_{\rm BGC}\sim2$ in the radial region where both GC subpopulations overlap.
Hence, from Eq. \ref{eq:sig_p} we have that $\sigma_p=\sigma_0R^{-\gamma}$. We can fit for $\gamma_{\rm RGC}$ and $\gamma_{\rm BGC}$ using the observed data beyond \Re. This yields $\sigma_{\rm 0,RGC}=225$ \kms, $\gamma_{\rm RGC}\sim0.01$, $\sigma_{\rm0,BGC}=285$ \kms\ and $\gamma_{\rm BGC}\sim 0.025$.

For RGCs, $\alpha+\gamma>3$ implies that for $\beta>0$ (radial anisotropy) the velocity dispersion decreases with both $\beta$ and $r$, while for BGCs,  $\alpha+\gamma<3$  implies that $\sigma_p$ has a larger normalisation factor for $\beta>0$, i.e.,  the presence of radial anisotropy would enhance the dispersion of the BGCs compared to the isotropic case, in contrast to the RGCs where it would lower the dispersion.

For BGCs, we conjecture that the kurtosis would have the same reversal behaviour with respect to the usual $R^{1/4}$ systems with $\alpha>3$, i.e. a negative kurtosis would imply radial anisotropy, while a positive kurtosis would imply tangential anisotropy. We are not aware of any work that has examined this issue for the kurtosis explicitly, but note that such a reversal has been observed, e.g., in the central regions of the Coma cluster 
\citep[][Fig. 3]{2003MNRAS.343..401L}. This will be explored in detail in the following dispersion--kurtosis analysis. 

We use Eq. \ref{eq:jeansrad} to derive suitable anisotropy parameters as required for compatibility of the RGC and BGC dispersion $\sigma_p$ with the same potential. Indeed, despite the well known degeneracy between anisotropy and potential for each tracer individually, the solutions are linked to a common potential. Thus, we expect the combination of tracers to help break the degeneracy. 

We first assume that the dynamics of the two subpopulations are influenced by the same total mass distribution as (i.e. $\sigma^2_r(r)$ is unique for both tracers in Eq. \ref{eq:jeansrad}).  Thus, solving and equating $\sigma^2_r(r)$ in Eq. \ref{eq:jeansrad} for both RGCs and BGCs at some fixed radius (e.g. $\overline{r}=300''$) we obtain:
\begin{eqnarray*}
\frac{\overline{r}}{G}\sigma_{\rm r,RGC}^2(\overline{r}) \,[\alpha_{\rm RGC}(\overline{r})+\gamma_{\rm RGC}(\overline{r})-2\beta_{\rm RGC}(\overline{r})]=
\\
\frac{\overline{r}}{G}\sigma_{\rm r,BGC}^2(\overline{r}) \,[\alpha_{\rm BGC}(\overline{r})+\gamma_{\rm BGC}(\overline{r})-2\beta_{\rm BGC}(\overline{r})].
\label{eq:jeansrad2}
\end{eqnarray*}
We can then solve for $\beta_{\rm BGC}$ by assuming some value for $\beta_{\rm RGC}$. 
This produces the following $\beta$ sample pairs:\\
~\\
$\{\beta_{\rm RGC}, \beta_{\rm BGC}\}=\{-1,-0.59\},\{-0.5,-0.32\},\{0,-0.05\},$

$\{0.20,0.07\},\{0.40,0.18\},\{0.6,0.30\},\{0.8,0.42\}$,\\
~\\
all of which are compatible with the observed $\sigma_p$ because they are algebraically derived through combining Eq. \ref{eq:jeansrad} and Eq. \ref{eq:jeansrad2}. Hence, we are faced with a further degeneracy.  The only way to break this degeneracy is by making use of the information provided by higher-order velocity moments (see following section). This simple algebraic derivation shows that significant differences in the intrinsic anisotropy parameter between both subpopulations are permitted by the observed dispersions in a single joint mass solution (i.e., both subpopulations in equilibrium within the same potential).

Excluding the nearly isotropic case (i.e., $\beta_{\rm RGC}=0, \beta_{\rm BGC}=-0.05$), it can be seen that tangential or radial anisotropies of the RGCs usually correspond to milder BGCs anisotropy biases in the same direction. In other words, the kinematics of the blue subpopulation are less sensitive to variations in the anisotropy.
 
\section{Dynamics: mass and anisotropy distribution}\label{sec:dynamics}

The dispersion--kurtosis procedure has been fully tested in ``round" elliptical galaxies both in the context of Newtonian potentials (see e.g. N+09, N+11) or in alternative gravity (N+12). This technique allows for the control of degeneracies between $\beta$ and the total mass, which is the sum of the stellar mass (see \S \ref{sec:data}, where the stellar $M/L$ is a free parameter) and the parametrised halo density profile. We assume that the latter follows an NFW distribution.

Under the assumptions of a spherical potential (see \S\ref{sec:data}), absence of rotation, and constant $\beta$ (corresponding to the $f(E,L)=f_0 L^{-2\beta}$ family of distribution functions, see \citealt{2002MNRAS.333..697L} and references therein), we follow N+12 and derive the 2-nd and 4-th moment radial equations (see also \citealt{1990AJ.....99.1548M}) compactly as:

\begin{equation}\label{eq:jeans2-4}
s(r)= r^{-2\beta}\int_r^\infty x^{2\beta} H(x) dx, \rm and
\end{equation}
\begin{equation} 
H(r)=\left\{ \rho\frac{G M_{\rm tot}}{r^2};
3\rho\frac{G M_{\rm tot}}{r^2}\overline{v_r^2} \right\}, 
\end{equation}
respectively, where $s(r)=\{\rho\sigma_r^2; \rho \overline{v_r^4}\}$ and $M_{\rm tot}$ is the sum of 
the stellar mass and a spherical dark halo.

The total stellar mass is given by $M_\star=M/L_\star \times L_{\rm tot}$, while the cumulative NFW halo mass is: 
\begin{equation}
M_{\rm d}(r)=4 \pi \rho_s r_s^3 A(r/r_s) 
\end{equation}
where $A(x) \equiv \ln (1+x)-x/(1+x)$. The two scale parameters $r_s$ (the scale radius) and $\rho_s$ (the characteristic density) are free parameters in the analysis. The fitted value of $r_s$ is typically of the order of some \Re\ and thus it is a parameter that can be directly derived using radially extended tracers such as GCs.

Alternatively, one may define the halo parameter space as the concentration ($c_{\rm vir}\equiv r_{\rm vir}/r_s$), and virial mass ($M\vir$) at the virial radius ($r\vir$). The latter is defined as the radius at which the mean halo density is $\sim100$ times the critical density $\rho_{\rm crit}=2.775\times10^{11} h^2 M_\odot$~Mpc$^{-3}$ (i.e. the virial overdensity value is $\Delta_{\rm vir}\simeq 97$ in WMAP7 cosmology, where we take $h=0.7$ as per \citealt{2011ApJ...740..102K}). However, the virial quantities are more difficult to constrain as they are not as readily accessible using the current kinematic data. As these quantities are typically predicted by cosmological simulations, we extrapolate them from our model to the theory. We compare our results with predictions from \LCDM\ cosmology simulations based on WMAP7 parameters (\citealt{2011ApJ...740..102K}). We also verify whether the $c\vir$ and $M\vir$ values are compatible with different dark matter flavours such as Warm Dark Matter (WDM, e.g. \citealt{Schneider12})\footnote{As we use a NFW density profile to fit the WDM haloes, we may compare our virial estimates with WDM simulations directly.}. 

Finally, we consider models including adiabatically contracted haloes in order to account for the effect of the baryon infall that may alter the halo density profile as predicted in collisionless simulations. For this, we use the standard recipe from \citet{Gnedin04}.

The projected velocity moments to be fitted to the observed kinematics are defined as:
\begin{equation}  
\label{eq4}
  \sigma_{\rm los}^2 (R) = \frac{2}{I(R)} \int_{R}^{\infty}
  \left( 1-\beta \frac{R^2}{r^2} \right) \frac{j_* \,
  \sigma_r^2 \,r}{\sqrt{r^2 - R^2}} \,{\rm d} r \ ,
\end{equation}
and 
\begin{equation}    
\label{eq7}
\overline{v_{\rm los}^4} (R) = \frac{2}{I(R)} \int_{R}^{\infty} \left(1 - 2 \beta \frac{R^2}{r^2} + \frac{\beta(1+\beta)}{2}
  \frac{R^4}{r^4}\right)
\frac{j_* \, \overline{v_r^4} \,r}{\sqrt{r^2 - R^2}} 
\,{\rm d}r ,
\end{equation}
with $I(R)$ and $j(r)$ the 2D and 3D surface densities of the tracer respectively. 
From these, we compute the projected kurtosis as $\kappa(r)={\overline{v_{\rm los}^4}}/\sigma_{\rm los}^4$.

Even though the equations listed above are valid for cases where $\beta=\rm const$, we now generalise their application. To explore a wide range of anisotropy profiles in the dispersion--kurtosis dynamical analysis, we adopt the following $\beta(r)$ parametrisation introduced by \citet{Churazov10}:
\begin{equation}\label{eq:betar}
\beta(r)=\frac{\beta_2 r^c+\beta_1 r_a^c}{r^c+r_a^c},
\end{equation}
which is characterised by two {\it plateaus} corresponding to $\beta_1$ and $\beta_2$ that set the asymptotic values for $r\rightarrow0$ and $r\rightarrow\infty$, respectively. Over these radial ranges we confirm that our assumption that $\beta$ is constant still holds. Therefore, we may use only the kurtosis values measured at the low and high extremes of our radial range to constrain the anisotropy parameters. In this case, the contribution from the $c$ parameter that determines the steepness of the transition from $\beta_1$ to $\beta_2$ is marginal, while $r_a$ is the scale at which this transition occurs.
For convenience, we fix $\beta_1=0$ (as previously measured by e.g. \citealt{Cappellari07}) and $r_a=200''$ as this is the radius where the kurtosis shows a sudden rise toward positive values (see Fig. 
\ref{fig:fig_kin}). At this radius, it is likely that the RGC anisotropy changes significantly. We also try varying $r_a$ and find that it does not have a significant effect on our results. For practical reasons and to reduce the number of free parameters, we also fix the slope parameter to $c=6$, as tests have shown that the current dataset cannot reliably constrain this parameter.

In the following sections, we model the observed dispersion and kurtosis in order to derive the best fit parameters of our models, which include the stellar mass-to-light ratio, the dark halo parameters ($r_s$ and $\rho_s$) and the asymptotic anisotropy ($\beta_2$).
Our approach is designed to solve the mass-anisotropy degeneracy mainly in the outer
regions in order to have unbiased estimates of the halo parameters.
We iteratively solve Eqs. 2, 3, 5 and 6 and minimise $\chi^2$ as defined by
\begin{equation}
\chi^2=\sum_{i=1}^{N_{\rm data}} \left[\lambda_i\frac{p^{\rm obs}_i-p^{\rm mod}_i}{\delta p^{\rm obs}_i}\right]^2 ,
\label{chi2}
\end{equation}
where $p^{\rm obs}_i$ are the observed data points ($\sigma_{\rm los}$ and $\kappa_{\rm los}$), $p^{\rm mod}_i$ the model values, $\delta p^{\rm obs}_i$ the uncertainties on the observed values, all of which are measured at radial position $R_i$. When combining the $\chi^2$ of the $\sigma_{\rm los}$ and $\kappa_{\rm los}$ to infer the model parameters consistent with the separate GC tracers, we apply a weight to each bin according to the penalisation factor $\lambda_i\sim1/N_{\rm data}$.

\begin{figure*}
\hspace{-0.5cm}
\vspace{-0.5cm}
\psfig{file=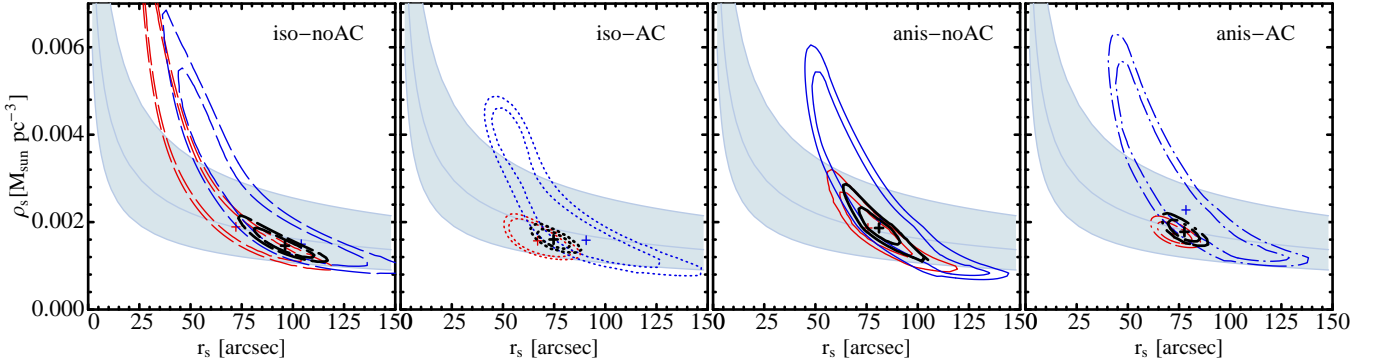,width=1.03\textwidth}
\caption{Confidence contours in the $\rho_s-r_s$ halo parameter space for the RGC (red curves) and BGC (blue curves) dynamics. Each panel reports a different model combination (see \S\ref{sec:self}). Best-fit halo parameters are given with red and blue crosses respectively. The combined confidence contours for each model are plotted as black lines with the best fit values marked with a black crosses. The $\chi^2$ for the best-fit combined solutions are given in Table 1. The filled area shows the expectation from cosmological simulations in the $\Lambda CDM$ and WMAP7 cosmology as per \citet{2011ApJ...740..102K} after converting $(c\vir-M\vir)$ to $(\rho_s-r_s)$.}\label{fig:fig_mods_1}
\end{figure*}

\begin{table*}
\caption{Summary of the best-fit multi-component model parameters. Typical uncertainties on the \Ystar\ , $\log {M_*}$ and $\beta$ parameter are of the order of 0.1\Ysol\, 0.05 dex and $\sim$0.1, respectively. For the $AC$ models, halo parameters are given pre-contraction, while $f_{\rm DM}$, $\Upsilon(\Re)$, $\Upsilon(5\Re)$ and $\Upsilon(R\vir)$ are given after contraction ($\Re=81''=9.1$ kpc). The expected $c_{\rm vir}$ from WMAP7 are indicated in squared brackets.}
\label{tab:jeanssumm}
\scriptsize
\noindent{\smallskip}\\
\hspace{-1.cm}
\begin{tabular}{lcccccccccccc}\hline\hline
Model  &  \Ystar$$& $\log {M_*}$ & $\rho_s/10^{-3}$ & $r_s$& $c_{\rm vir}$ & log ${M\vir}$& $f_{\rm DM}(5\Re)$ & $\beta(5\Re)$ &$\Upsilon(\Re)$ & $\Upsilon(5\Re)$ & $\Upsilon(R\vir)$ & $\chi^2$/d.o.f. \\
 &  (\Ysol)  & (\Msun) &  (\Msun/pc$^3$) & (kpc) &  & (\Msun) & & RGC/BGC &(\Ysol)  & (\Ysol)  & (\Ysol)  &  (\#par) \\
\hline
iso-noAC & $8.2$ & 11.90& 1.5$^{+0.4}_{-0.3}$ & 96$^{+19}_{-20}$ & 7$^{+6}_{-5}$[8] & 13.33$^{+0.30}_{-0.34}$ & 0.71$^{+0.10}_{-0.13}$ & 0/0 & 11$^{+2}_{-1}$ & 28$^{+13}_{-8}$ &217 & 16/22(3)\\
iso-AC&  $7.0$ & 11.84 & 1.6$^{+0.3}_{-0.2}$ & 75$^{+15}_{-13}$ & 8$^{+9}_{-4}$[8] &  13.06$^{+0.25}_{-0.25}$ & 0.71$^{+0.09}_{-0.10}$ & 0/0 & 13$^{+1}_{-1}$ & 24$^{+8}_{-5}$ &105 & 18/22(3) \\
ani-noAC  & $8.0$ & 11.89 &1.9 $^{+0.4}_{-0.2}$ & 81$^{+4}_{-16}$& 8$^{+5}_{-4}$[8] & 13.24$^{+0.14}_{-0.28}$  & 0.71$^{+0.05}_{-0.10}$ & 0.43/0.15 & 11$^{+1}_{-1}$ & 28$^{+6}_{-7}$ & 178 & 12/18(4)\\
ani-AC  & $6.0$ & 11.77& 1.8 $^{+0.2}_{-0.2}$ & 78$^{+12}_{-11}$& 8$^{+9}_{-5}$[8] & 13.16$^{+0.19}_{-0.22}$  & 0.74$^{+0.07}_{-0.08}$ & 0.4/0.2 & 13$^{+1}_{-1}$ & 26$^{+6}_{-5}$ & 135 & 23/18(4)\\
\hline
\hline
\end{tabular}
\end{table*}

\section{Combining Red and Blue globular cluster dynamics}\label{sec:models}

The combination of multiple dynamical tracers is the best way to break all degeneracies associated with dynamical modelling of hot systems. Because different GC subpopulations usually have decoupled kinematics that can be probed with a single observational set-up, they constitute a natural and powerful tool for degeneracy-free mass modelling. 

\subsection{Best-fit halo parameter determination}
We have shown that the RGCs in NGC~5846 reliably trace the kinematics of the galaxy stars. Hence, we combine the integrated light data with those of the RGCs (surface density and kinematics) into a single tracer family. The BGCs being a second set of spatially and kinematically decoupled tracers. The final two datasets are of different quality in terms of their spatial sampling and error budget (Fig. \ref{fig:fig_kin}). Therefore, we adopt the following approach to combine constraints from RGCs and BGCs:

1) we model the innermost stellar dispersion and kurtosis together with that of the RGC  to obtain a full set of stellar and halo parameters (\Ystar, $\rho_s$, $r_s$, $\beta$), where $\Ystar$ is the dynamically derived stellar $\ML$. The high quality central kinematic data (mainly long slit) constrain the central $M/L$ and anisotropy more efficiently, while the extended RGC kinematics help constrain the halo parameters and large radii anisotropy for RGCs.

2) Once \Ystar\ in the total potential is fixed by modelling the integrated light and RGCs, we use the BGC dispersion and kurtosis independently to constrain the dark halo parameters.

3)  We find the system's self-consistent halo model where both RGC and BGC halo solutions overlap. The confidence intervals of the self-consistent halo model are obtained from the sum of both $\chi^2$ distributions with best fit parameters corresponding to the minimum in combined $\chi^2$.  Separate solutions for other parameters (\Ystar, $\beta$) are obtained for each tracer set.

4) We perform the analysis assuming either a standard NFW or an adiabatically contracted NFW halo ($noAC$ and $AC$, respectively), as well as either the isotropic case (i.e. $\beta=0$) or a radial anisotropic profile $\beta(r)$ as per Eq. \ref{eq:betar} ({\it iso} and {\it ani}, respectively).


\begin{figure*}
\hspace{-0.1cm}
\vspace{-0.5cm}
\psfig{file=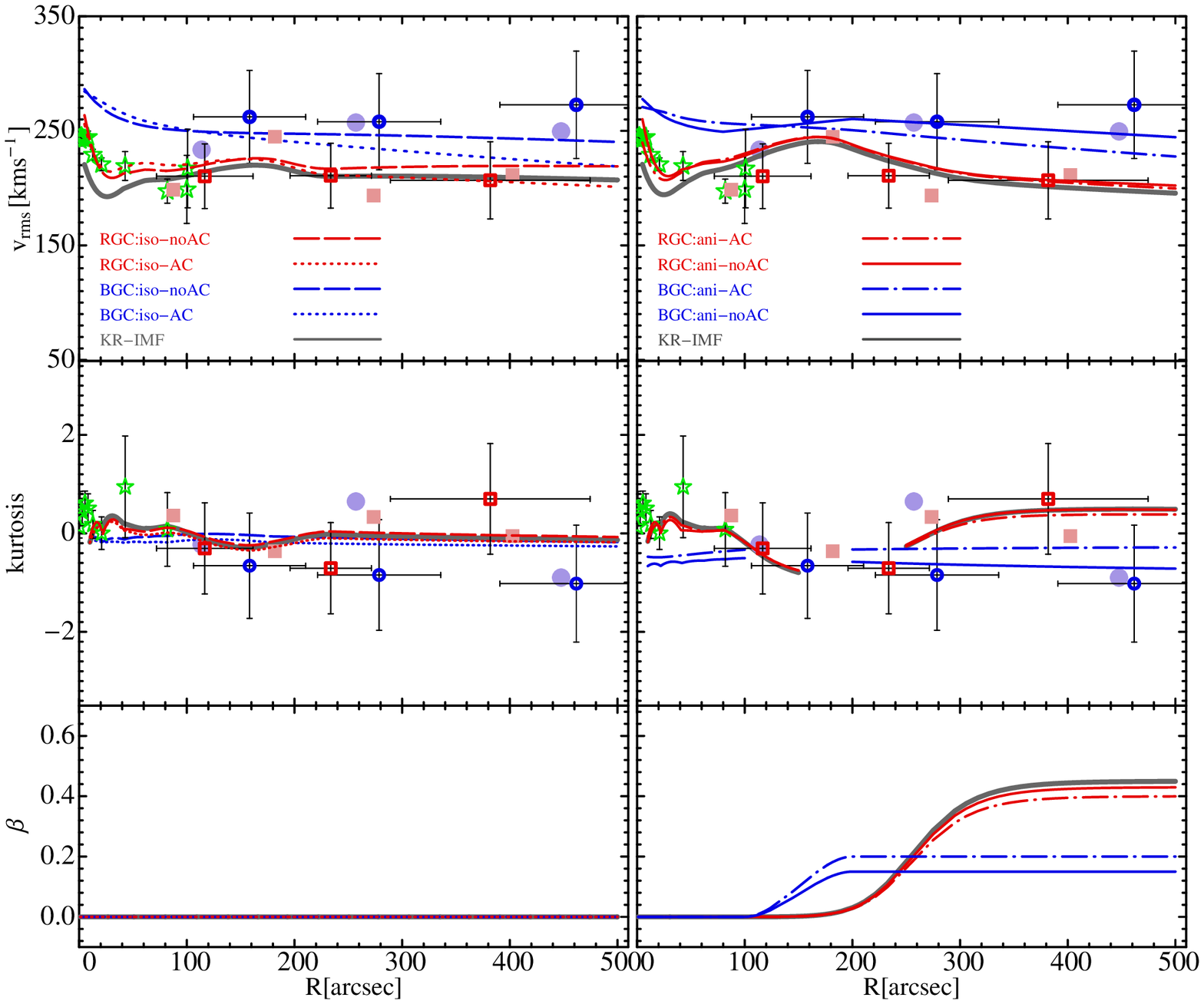,width=0.8\textwidth}
\caption{Joint best fit models of the dispersion (top), kurtosis (central) and anisotropy (bottom) profiles of RGC (red
curves) and the BGC (blue curves). Symbols follow Fig. \ref{fig:fig_kin} and curve styles follow Fig. \ref{fig:fig_mods_1}. The left panels show two isotropic models with $noAC$ and $AC$ haloes. The grey line is the forced solution for the RGC when converting to a Kroupa IMF with an $AC$ halo (see \S\ref{sec:summary}). The right panels shows the best fit red and blue anisotropic $noAC$ and $AC$ models as labeled. Also shown in grey is the best-fit $AC$ halo with a Kroupa IMF and orbital anisotropy.}\label{fig:fig_mods_2}
\end{figure*}

Best fit parameters for the various model combinations described in item 4 above are reported for the joint solution of the RGCs and BGCs. Individual red and blue (items 1 and 2 above) confidence contours of the halo parameters ($\rho_s$, $r_s$) marginalised over other fitted parameters (\Ystar, $\beta$) are shown in Fig. \ref{fig:fig_mods_1}.  Also plotted in black in the same figure are the combined (item 3 above) $\chi^2$ for the joint halo best fit parameters, along with the expectation from cosmological simulations in the $\Lambda CDM$ and WMAP7 cosmology as per \citet{2011ApJ...740..102K} with $c\vir-M\vir$ converted to $\rho_s-r_s$. As direct predictions of these quantities within $\Lambda CDM$ simulations are not yet available, our derived expectations are obtained in a manner similar to that used for independent indirect inferences (e.g. \citealt{2012MNRAS.423.2177S}, Eq. 3).

Fig. \ref{fig:fig_mods_2} shows the  joint solutions for the various models. As we are interested in the joint solutions, we do not discuss the independent RGC and BGC solutions (red and blue contours in Fig. \ref{fig:fig_mods_1}) further. There is no additional information contained in these solutions compared to the joint solutions except for the fact that they are optimised for the individual tracer kinematics. Hence, they generally have lower $\chi^2$ than the combined best-fit and better reproduce the individual kinematic profiles than the joint models. 

\subsection{Self-consistent models}\label{sec:self}

We now closely inspect the best-fit models for RGCs and BGCs with a particular focus on the joint solutions that have been obtained by combining their $\chi^2$ distributions as described in the previous section. 

\subsubsection{Isotropic case: no-AC and AC models}
Starting from the simple assumption of orbital isotropy (i.e. $\beta=0$), \Ystar\ is allowed to vary from $4.5$ to $9.5~\Ysol$, a range that encompasses the $M/L$ predictions of both Chabrier and a super-Salpeter IMF. 

For the $iso$-$noAC$ case, $\chi^2$ in halo parameter space is minimised using \Ystar$=8.2\Ysol$, which is consistent with a Salpeter IMF.  The resulting individual confidence contours for the RGC and BGC models (Fig. \ref{fig:fig_mods_1}) are stretched across the parameter space, which illustrates the strong degeneracy of the halo parameters.  The situation is much improved for the $iso$-$AC$ case where the confidence contours of the RGCs are more compact, and hence the degeneracy is somewhat alleviated. The best fit stellar $M/L$ for $iso$-$AC$ is \Ystar$=7.0\Ysol$, which is close to the lowest limit allowed by stellar populations with a Salpeter IMF.

In both cases the confidence contours overlap and converge to a combined solution. The parameters that minimise the $\chi^2$ are reported in Table 1. Overall, the halo parameters for the two isotropic models are consistent with each other. Even though the best fit $iso$-$noAC$ model has a slightly larger density ($\rho_s$) and concentration parameter ($c\vir$), it is fully consistent (i.e. within the confidence area) with the $iso$-$AC$ model.  

In the left side of Fig. \ref{fig:fig_mods_2}, the dispersion curve of the RGCs and BGCs predicted by the combined solutions to the isotropic models are well fitted for both the $iso$-$noAC$ and $AC$ cases. The former performs somewhat better for the BGCs. On the other hand, the fit to the kurtosis is quite poor outside $\sim150''$ (i.e. $\sim 2 \Re$) for both subpopulations. In particular, the modelled RGC kurtosis systematically underpredicts the observed one beyond $R>200''$, suggesting the presence of radially biased orbits. In contrast, the modelled BGC kurtosis slightly overpredicts the measured one, possibly indicating the presence of some level of anisotropy. In the central regions 
our working hypothesis is that the orbits in the galaxy core are basically isotropic for both subpopulations. 

Overall, the poor fit of the kurtosis profile at large radii does not seem to strongly affect the global significance of the final fit, which remains fairly good for both cases. 

\subsubsection{Anisotropic case: no-AC models}
The introduction of the parametrised $\beta(r)$ profile in Eq. \ref{eq:betar} allows a considerably better match of the model with the outer galaxy regions. The $ani$-$noAC$ fit to the RGCs kinematics has minimum $\chi^2=9.6/12$ for $\Ystar=8$, which is fully compatible with a Salpeter IMF and is in line with that found in the isotropic case.  The confidence contours of the halo parameters for this solution ($ani$-$noAC$) are plotted in Fig. \ref{fig:fig_mods_1}. The joint solution is also shown and agrees nicely with the WMAP7 prediction area.

The narrower range of halo parameter space for the $ani$-$noAC$ versus the $iso$ models is not a reflection of the relative $\chi^2$ values that are, in fact, fairly similar, once normalised to the number of degrees of freedom. Rather it is a result of the ability of the former to better reproduce the overall data in Fig. \ref{fig:fig_mods_2}. The dispersion curves of the two models all match the observed velocity dispersion profile of the RGCs, including the ``bump'' around $R\sim180''$ found for the $g-i=0.95$ colour cut. We confirm that this bump is real despite being hidden in the improved colour selection sample.  Using four bins with $\sim22$ RGCs, instead of the adopted 3 bins with $\sim 28$ RGCs, we also find a high dispersion of $\sim250$ \kms\ around $R\sim180''$. Moreover, the kurtosis profiles of the RGCs match the observations at all radii including the two {\it plateaux} in the anisotropy profile. In particular, as reported in the lower panel of Fig. \ref{fig:fig_mods_2}, the $ani$-$noAC$ solution of RGCs shows a moderate ($\beta\sim0.4$) radial anisotropy corresponding to the positive values of the kurtosis profile in the outermost regions.

The BGC kinematics are best fitted with $\beta\sim0$ in the central regions and mild radial anisotropy  ($\beta\sim 0.15$) outside $r_a=100''$. In this case, the $ani$-$noAC$ is a remarkably better match to the observed dispersion and kurtosis than the $iso$ model at all radii. It is interesting that the joint solution robustly shows a difference in the orbital distributions of the two subpopulations as mainly imprinted in the kurtosis profiles. Indeed, the intricate combination of the density and the velocity dispersion slopes (\S\ref{sec:effect}) conspire to produce significantly different kurtosis profiles for similar asymptotic radial anisotropies. The dispersion--kurtosis analysis is sufficiently sensitive to highlight this difference. 

\subsubsection{Anisotropic case: AC models}

We also fit the adiabatic contraction model for the anisotropic case. Similarly to the isotropic case, we obtain a drastic reduction of the confidence area in halo parameter space around the models. The converging solution uses $\Ystar=6\Ysol$, similar to the Kroupa IMF, and a halo model whose best fit parameters lie between the $ani$-$noAC$ models (see Table 1) with which they are consistent within $1\sigma$ (see Fig. \ref{fig:fig_mods_1}). In Fig. \ref{fig:fig_mods_2}, the dispersion and kurtosis profiles of the RGCs are nearly indistinguishable from the $ani$-$noAC$ case with an overall slightly larger $\chi^2$. In particular, the $ani$-$AC$ model fails to reproduce the central RGC dispersion--kurtosis. Also in Fig. \ref{fig:fig_mods_2}, the best fit to the BGCs involves a degree of anisotropy that is slightly larger than for the $noAC$. In this case, the lower velocity dispersion normalisation compared to the $ani$-$noAC$ case (right-top panel) drives the kurtosis normalisation to larger values (we recall that $\kappa=\overline{v_p^4}/\sigma_p^4$).

On the basis of the overall poorer significance and, in particular, the worse fit to the BGCs, the $ani$-$AC$ may be ruled out. This has consequences for the range of possible $\Ystar$, as values consistent with a Kroupa IMF are only accommodated when considering some standard $AC$ recipe (\citealt{Gnedin04}).
 
In order to make this conclusion more convincing, we force the $AC$ models (both isotropic and anisotropic) of RGCs to a nominal Kroupa IMF, $\Ystar=5\Ysol$ (see Fig. \ref{fig:fig_mods_2}) and still find that it is possible to find a good fit to the data at large radii. However, the match to the central data is unacceptably poorer than for higher $\Ystar$. The $\chi^2$ is about twice as large as the corresponding Salpeter-like IMF solutions, thus we can confidently rule out bottom light IMFs (see also below) at the $>2\sigma$ level. We do not explore stronger $AC$ recipes (e.g., \citealt{1986ApJ...301...27B}), which might allow lower stellar mass normalisation by dragging more DM into the central regions (see e.g. \citealt{Napolitano10}), as these seem to be disfavoured by both simulations (e.g. \citealt{2010MNRAS.407..435A}) and observational studies (e.g., \citealt{Auger10}).

\section{Discussion}\label{sec:summary} 
\subsection{Orbital anisotropies and implications for formation processes}
On the basis of the results presented above, we rule out the $iso$ models because of their failure to reproduce the kurtosis profiles. In contrast, the $ani$ models allow us to fit both the dispersion and the kurtosis profiles, in particular in the outer regions. The implied difference in the anisotropy distributions of the RGC and BGC subpopulations is statistically significant and matches the predictions from \S\ref{sec:effect} (i.e. $\beta_{\rm RGC}=0.4$, $\beta_{\rm BGC}=0.18$).

Thus, the joint dispersion--kurtosis analysis of RGCs and BGCs allows us to break the intrinsic degeneracy (as discussed in \S\ref{sec:effect}) of the two subpopulations anisotropy under the single equilibrium potential. This is a remarkable result as direct estimates of the anisotropy of GC  subpopulations in individual galaxies have so far been based on simplified approaches (\citealt{Romanowsky09}) or circumstantial evidence (\citealt{Schuberth12}; P+13), rather than on direct modelling of higher-order velocity moments. 

The final orbital anisotropy is thus in agreement with the expectations from N-body simulations, which predict that DM halos and their baryonic tracers are radially anisotropic in their outer regions owing to infall and merger processes (\citealt{Dekel05,2005MNRAS.363..705M,2006NewA...11..333H,2006MNRAS.370..681N,2007MNRAS.376...39O}). In particular, RGCs show a degree of radial anisotropy that is consistent with model predictions (\citealt{Dekel05}) and previous estimates from planetary nebulae kinematics in intermediate luminosity systems (N+09, \citealt{2009MNRAS.395...76D}). {\it Due to the established link between planetary nebulae and RGCs with field stars, this shared dynamical behaviour likely descends from a common assembly process.}

On the other hand, our finding of a distinct orbital distribution for the BGCs may be the ``smoking gun" of a different origin for this subpopulation. The dynamical proof of this hypothesis is still lacking (see \citealt{Romanowsky09} for a discussion). In the context of the multi-phase galaxy formation scenario (\citealt{1997AJ....113.1652F}), blue, more metal-poor GCs may form around their progenitor systems before the formation and assembly of the bulk of stars and associated red, more metal-rich GCs. Although age-dating of extragalactic GCs can only distinguish the ages of BGCs and RGCs to within $\sim$ 2 Gyrs (\citealt{2005AJ....130.1315S}), recent work on Milky Way GCs seems to confirm that the BGCs are systematically older than the RGCs by $\sim$ 0.8 Gyrs (\citealt{2011ApJ...738...74D}).

In the alternative \citet{1998ApJ...501..554C} scenario, the RGCs also should mimic the starlight, while BGCs form at early times but are accreted from (radially) infalling satellites. Hence, the orbital distribution of BGCs should reflect the earlier formation, while the RGCs should follow that of the stars. We have seen that this is usually the case for the RGC as confirmed through their matching dynamics to those of other stellar tracers (e.g. long slit data and planetary nebulae).

The earlier phase assembly of the metal poor GCs in the context of a full cosmological picture of galaxy assembly has been investigated by \citet{2008ApJ...689..919P}, who found that 
the metal-poor GC orbits are set by the orbits of their progenitor satellite galaxies, resulting in near-isotropy ($\beta\sim\pm0.2$) out to $\sim0.1$ virial radii and radially biased outward orbits (out to $\beta\sim 0.5$). This is qualitatively consistent with a centrally isotropic $\beta$ profile transitioning to mild radial anisotropy ($\beta\lsim0.2$) outside $\sim 2\Re$, which allows us to fit the dispersion and kurtosis of BGCs with very good accuracy.

\subsection{Dark matter halo properties and IMF}\label{sec:IMF}

The overall picture emerging from our results summarised in Table \ref{tab:jeanssumm} is that the overall joint solutions are strikingly stable despite having different halo solutions associated with the considerably varied anisotropy and halo contraction assumptions (see Fig. \ref{fig:fig_mods_1}) in our model procedure. For the $noAC$ models, the concentration is $c\vir=7-8$ and virial mass $\log M\vir/\Msun\sim13.3$, while the $AC$ solutions also give $c\vir=8$ with a slightly smaller virial mass $\langle \log M\vir/\Msun\rangle=13.1$.

Generally, the joint solutions agree with the WMAP7 based \LCDM\ predictions both in terms of $c\vir-M\vir$ (\citealt{2011ApJ...740..102K}) and corresponding $\rho_s-r_s$ (Fig. \ref{fig:fig_mods_1}). This is true despite the separate halo solutions for the RGCs and BGCs ranging over large areas of parameter space (Fig. \ref{fig:fig_mods_1}), although remaining within $1\sigma$ of the WMAP7 expectation. 

The $c\vir-M\vir$ pairs from Table 1 are also nicely consistent with most of the WDM model predictions from \citet[][e.g., their Fig. 11]{Schneider12}. In this model, a concentration parameter of $c\vir\sim8$ is expected at virial masses $\log M\vir/\Msun=13.1-13.3$ for $m=0.5-1$ keV WDM particles, while at the same masses $c\vir\sim7$ is predicted for $m=0.25$ keV. Unfortunately, at the mass scales of massive ellipticals, the WDM predictions do not differ from the \LCDM\ $c\vir-M\vir$ and it is not possible to disentangle the different cosmologies. However, it is encouraging that the tight constraints on the halo parameters obtained using the RGC/BGC model combination in lower mass systems will allow discrimination 
of one DM flavour over the other. 
   
As previously mentioned, the narrow distribution of virial parameters seems insensitive to the quite large changes of the anisotropy parameter, which varies from $\beta=0$ to $\beta\sim0.4~(0.2)$ for the RGCs (BGCs, see Table 1). This is an interesting feature emerging from our analysis and implies that for massive systems with fairly flat dispersion profiles, the constraints on the halo parameters are almost independent of the anisotropy assumption (see also the PN analysis of NGC 4374 in N+11, their Table 2), and the overall anisotropy can only be constrained through the higher velocity moments (i.e. the kurtosis in our case). The main advantage of using different tracers such as the two GC subpopulations is significantly reduced uncertainties in the measured halo solutions as compared to those yielded by single extended tracer (i.e. PNe) as in N+11. For instance, uncertainties in $c\vir$ are reduced by up to a factor of $\sim4$.

The assumption on halo contraction evidently affects the final stellar $M/L$ inferred to fit the galaxy kinematics, especially in the inner regions. We find a mean $\langle M/L \rangle\sim 6.5\Ysol$ slightly lower than predicted by stellar population models using a Salpeter IMF for the $AC$ models (see \S\ref{sec:intro}), that is $\sim$ 20\% lower than that obtained for the $noAC$ cases ($\langle M/L \rangle\sim 8\Ysol$), which is consistent with a Salpeter IMF. We also show that bottom-lighter IMFs (Chabrier or Kroupa IMF) are confidently ruled out, even when allowing for halo contraction. {\it This is independent confirmation of recent findings that stars in high mass galaxies form with a bottom-heavier IMF} \citep{2012ApJ...760...71C,2012Natur.484..485C,2012AJ....144...78W,Tortora13,2013MNRAS.tmp.1627L}. To our knowledge, this work is the first study using a radially extended dataset to constrain the IMF along with an accurate determination of the halo parameters.

\begin{figure}
\hspace{-0.1cm}
\psfig{file=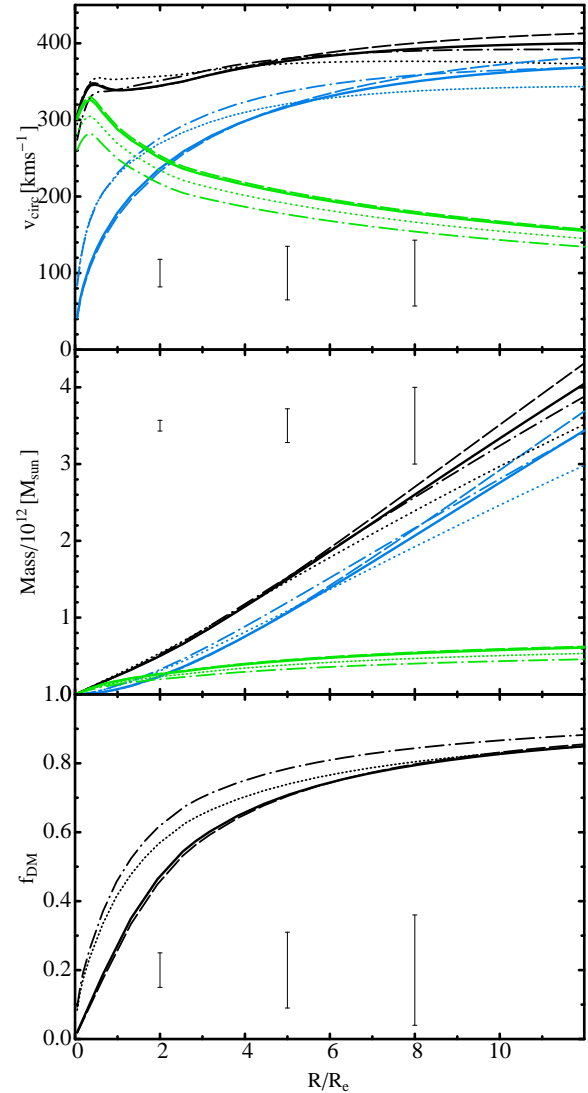,width=0.46\textwidth}
\caption{Summary of the galaxy mass distribution. {\it Upper panel:} circular velocity of the stellar mass (green), dark mass (cyan) and total mass (black) for models shown in Fig. \ref{fig:fig_mods_2} with identical curve styles. {\it Middle panel:} same as above for the stellar and total mass distribution. {\it Lower panel:} dark matter fraction ($M_{\rm DM}/M_{\rm tot}$). Typical uncertainties are plotted for various radii}
\label{fig:fig_mods_3}
\end{figure}

The degeneracy between $\Upsilon_\star$ and halo contraction is broken thanks to the large radius baseline covered by the stellar and RGC data.  The stellar data constrain the total mass within $\sim R_e$, where the central DM contribution follows from the adopted $\Upsilon_\star$ value. The DM profile at large radii ($\sim$~1--5~$R_e$) differs depending on whether it is contracted or not, an ambivalence that is constrained by the RGCs. The BGC data further tighten the constraints on the halo parameters.

Although we have not directly explored DM halos with shallow central profiles (e.g. \citealt{2009MNRAS.397.1169D}), we can qualitatively infer that such halos would push more of the central mass into the stellar component, thus requiring a 
super-Salpeter IMF.

Parametrizing IMF variations using the $\delta_{\rm IMF}$ parameter as a measure the departure from the typical Milky Way IMF, we obtain $\delta_{\rm IMF}=1.8\pm0.2$ for NGC 5846, which is consistent with the typical values found in SDSS galaxies and other literature data at comparable stellar masses \citep{Tortora13}.

\subsection{Comparison with other works}\label{sec:discuss}

The giant elliptical NGC 5846 is a dark matter dominated system with very high virial $M/L$ ($\Upsilon(R\vir)>100\Ysol$, see Table 1). The circular velocity ($v_{\rm circ}$) profiles of the models discussed herein exhibit the typical features of a massive dark matter halo (Fig. \ref{fig:fig_mods_3}). For all models and after reaching a local minimum at $\sim\Re$, the $v_{\rm circ}$ increases around $R>3\Re$ before flattening out at the typical $r_s$ scale (i.e. $8-10$ \Re, see Table 1). 

The total mass distributions of the various models are all nearly linearly increasing with radius as shown in the middle panel of Fig. \ref{fig:fig_mods_3}. In the same figure, the corresponding DM fractions (lower panel) are plotted. These show that the DM mass exceeds that of the stars ($f_{\rm DM}>0.5$ by definition) at $1 \Re$ for the contracted models and $\sim 2\Re$ for the non contracted ones.

The DM fractions within $1\Re$ are of order $f_{\rm DM}\sim0.25$ for the normal DM solutions and $f_{\rm DM}\sim0.4$ for the contracted halo ones. This result is consistent with the typical central DM fractions measured with large statistical sample analyses from either galaxy dynamics and virial analysis \citep{Tortora09,Thomas09,Napolitano10,Grillo10}, or lensing studies \citep{Auger10,Tortora10} using a Salpeter IMF.

In all cases presented in Fig. \ref{fig:fig_mods_3}, typical uncertainties in derived dynamical quantities are plotted. As pointed out in \S\ref{sec:IMF}, it is clear that all models are generally consistent within typical uncertainties. This reflects the stability of the virial solution against different assumptions of anisotropy that are compatible with the observed kinematics.

The dynamical $M/L$ within \Re\ ($\Ystar=11-13$, see Table 1) is consistent with previous analyses from the ATLAS$^{\rm 3D}$ project \citep{Cappellari13}, wherein a value of $M/L_r=8.1$ was found. Once converted to the $V-$band (see \S\ref{sec:ML}), this corresponds to $M/L_V\sim12.5$. Our $M/L$ estimate is lower than K+00 (i.e. $M/L_B\sim11$ at \Re\ ), once the latter is converted to the $V-$band and corrected to a common distance. This may be a consequence of their assumption of a cored halo, which tends to show a flatter $M/L(r)$ than the cuspy NFW profiles used in this work. 

\begin{figure}
\hspace{-0.4cm}
\psfig{file=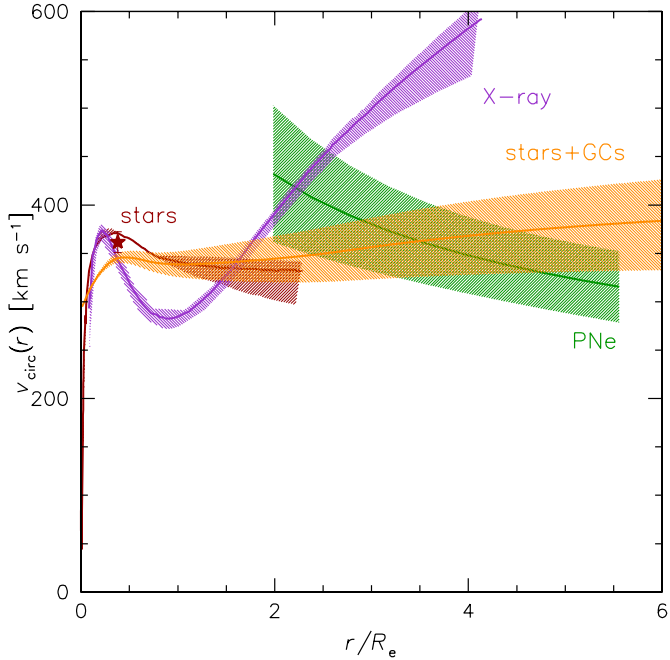,width=0.52\textwidth}
\caption{Comparison with previous analyses. The GC fiducial model from this work (stars+GCs, orange curve) is compared with previous star--only inferences (K+00, red curve, and ATLAS$^{\rm 3D}$, red star), a PN model from \citet{DeasonB}, and with an X-ray model from \citet{Das}. See text for a detailed discussion.}
\label{fig:vcirc_comp}
\end{figure}

However, in the central region, our $v_{\rm circ}$ models are dominated by the stellar component that peaks around 0.5 \Re\ with $\sim330-350\kms$. This is almost consistent with the findings of ATLAS$^{\rm 3D}$ (361\kms, \citealt{2013MNRAS.432.1862C}), but lower than the K+00 model solution. This is shown in Fig. \ref{fig:vcirc_comp}, where we see that we agree with K+00
outside $1\Re$ despite the smaller radial coverage of their star--only model, which does not allow comparison of the results at large radii. The central discrepancies between this and the previous two studies reflect the slightly larger stellar mass normalisation, compared to our estimate, that  they both inferred (i.e. $\Ystar\sim9.5$ and $\Ystar\sim9.2$, respectively, for K+00 and ATLAS$^{\rm 3D}$ at the same galactocentric distance and converted to $V-$band). This may be a consequence of the poor constraints on the overall dark halo derived from  their radially confined kinematics.

The study of \citet{DeasonB} is the one more closely related to our work in terms of data and model extent (see Fig. 7). Their analysis differs in that they do not fit the stellar component, but instead provide the solutions for both a Chabrier and a Salpeter IMF. In our analysis, we resolve this ambiguity by explicitly fitting for the IMFs and find that a heavier stellar mass normalisation is favoured over Chabrier IMF.

Comparing our results to the \citet{DeasonB} solution with a Salpeter IMF,  we only find agreement for $f_{\rm DM}(5\Re)$. This is a consequence of their smaller total stellar mass, which may be due to the limited radial extent of their selected surface brightness profile compared to that used here. 

In Fig. \ref{fig:vcirc_comp}, the \citet{DeasonB} $v_{\rm circ}$ profile is steeper and higher than all other models within 2.5\Re\ . This may reflect their adoption of a constant slope to describe the tracer density profile that is not a good representation of the central galaxy regions.
On larger scales, their $v_{\rm circ}$ also shows a steep decline that is significantly tilted with respect to our NFW model based profile. This implies a slightly lower total mass. For example, our total mass estimate ($\sim1.6\times10^{12}\Msun$) is $\sim30\%$ larger than theirs at 5\Re. Furthermore, the PNe dispersion profile is slightly lower at all radii (see e.g. P+13), which they model using an average anisotropy of $\overline{\beta}=0.2$. This value is lower than that of the RGCs presented herein. Both of the above push their model toward lower values for the overall mass.
 
Finally, Fig. \ref{fig:vcirc_comp} also includes the X-ray model of \citet{Das}. Their circular velocity is consistent with the central region estimates from dynamical studies, but it diverges from $2\Re$ onwards. It is possible that the disturbed X-ray structure in this system (\citealt{2011ApJ...743...15M}) is not suitable for equilibrium analysis. 
 
\section{Conclusions}\label{sec:conclus}

We have presented the first self-consistent Jeans model analysis of red and blue GC subpopulations including a dispersion--kurtosis analysis to break the degeneracies between dark matter, anisotropy and IMF applied to the giant galaxy NGC 5846. 

The two GC subpopulations are accurately separated using their colour distribution and careful evaluation of their colour mix as a function of radius. A conservative separation allows us to improve the kinematics of the RGCs and BGCs. These turn out to be decoupled, with the velocity dispersion profile of the RGCs flattening to $210\kms$ outside 3\Re\ ($\Re=81''$), and the velocity dispersion profile of the BGCs also flattening, but being $\sim40\kms$ larger than that of the RGCs. The kurtosis ($\kappa$) profiles of the two subpopulations are fairly similar within $R\simeq3\Re$, where they both show $\kappa\sim -0.5$, which is consistent with the outermost stellar data points from long slit spectroscopy.
At larger radii they diverge with RGCs increasing toward $\kappa\sim 1$ and BGC gently decreasing toward $\kappa\sim-1$. The kurtosis values for the two subpopulations differ at the $\sim 2\sigma$ level at the largest radius probed.

We have modelled the kinematic data assuming a standard NFW profile, including the effect of adiabatic contraction \citep{Gnedin04} and allowing the anisotropy parameter to vary with radius up to a constant value,  after confirming that orbital isotropy, $\beta=0$, in the centre (see also \citealt{Cappellari07}) provides a very good match to the dispersion and kurtosis data for both RGCs and BGCs. The free parameters of our model are the dynamical stellar mass-to-light ratio (\Ystar) 
the NFW density scale ($\rho_s$) and characteristic radius ($r_s$), and the outer anisotropy value $\beta$. In particular, we allow the \Ystar\ to encompass a wide range of values in order to map the predictions of a number of (single slope) IMFs, from bottom light (e.g. Chabrier or Kroupa IMF) to bottom heavy (e.g. Salpeter IMF).\\
~\\
Our main results are:
\begin{itemize}

\item \Ystar\ is 8.2$\Ysol$ for the isotropic case and $8\Ysol$ for the radially varying anisotropy parameter $\beta(r)$ (fully consistent with a Salpeter IMF). Taking into account adiabatic contraction, the stellar $M/L$ is $\Ystar\sim6-7\Ysol$, possibly sub-Salpeter. A Kroupa IMF is ruled out as it provides a much poorer fit to the central velocity dispersion values. If we use the IMF mismatch parameter with respect to the Milky Way, we obtain $\delta_{\rm IMF}=1.8$, consistent with typical values obtained for massive ellipticals (see \citealt{Tortora+13}). 
   
\item We have performed a separate fit to the RGC and BGC kinematics (dispersion and kurtosis) and defined the joint solution for the two subpopulations by combining the $\chi^2$ distributions. These self-consistent solutions show the existence of a common halo compatible with the observed kinematics. The two subpopulations, though, show a different orbital distribution with the RGCs being more radially biased in the outer regions than the BGCs. The full isotropic solutions are generally less significant than the anisotropic cases. 

In Table 1 we show the best--fit models obtained for the galaxy. The overall halo parameters are rather stable, independently of the anisotropy with either an average concentration of $c\vir\sim8$ and virial (stellar and dark) mass 
$\log M\vir\sim13.3$ for the $noAC$ models, or $c\vir\sim8$ and $\log M\vir\sim13.1$ for the $AC$ cases. The favoured solution is the $ani$-$noAC$ with anisotropy parameter for RGC of $\beta_{\rm RGC}\sim0.4$ and for BGCs varying between $\beta_{\rm RGC}\sim0.15$ at $R>3\Re$ and isotropic at $R\lsim1.5\Re$. All parameters are fully compatible with $\LCDM$ expectations and WMAP7 (e.g. from \citealt{2011ApJ...740..102K}), but also with WDM simulations (e.g. \citealt{Schneider12}), since $\LCDM$ and WDM are quite similar at these mass ranges.

\item The difference in the orbital anisotropy between the two GC subpopulations is statistically significant and suggests a difference in the red and blue GC formation mechanisms. In particular the milder anisotropy of the BGCs is compatible with the results of dark matter only N-body cosmological simulations (\citealt{2008ApJ...689..919P}) following the assembly in present day galaxies of these systems through mergers of dwarf-like progenitors formed at high-redshift ($z\gsim3$). The larger radial anisotropy found in RGCs, is instead consistent with expectations from simulations and model predictions (\citealt{Dekel05,2005MNRAS.363..705M}) and arise from infall and merger processes.  Previous estimates of planetary nebulae in intermediate luminosity systems (N+09, \citealt{2009MNRAS.395...76D}) reinforce the emerging 
association between the RGC subpopulations and the early-type galaxy underlying stellar population (see e.g. \citealt{Forbes12}). This decoupled orbital distribution of the two GC subpopulations is consistent with a two--phase assembly scenario of early-type systems (\citealt{1997AJ....113.1652F}; \citealt{1998ApJ...501..554C}).

\end{itemize}

All the above conclusions implicitly assume that the two GC subpopulations are under equilibrium within a common halo. In systems where this condition is not satisfied, this may be the primary reason for the failure to find common solutions between RGCs and BGCs (see e.g., NGC 1399, \citealt{Schuberth}). 

There are two major novelties explored in this paper that 
exploit the use of GCs to push beyond other techniques usually adopted for the extended dynamics of early-type galaxies (e.g. PNe, \citealt{2004ApJ...602..685P,Napolitano,Napolitano11}, galaxy satellites, \citealt{2003ApJ...598..260P}).

The first is that GCs naturally provide two decoupled families of tracers under a single observational set-up and conditions, which is necessary to break the mass--anisotropy degeneracy.
The second is that the combination of RGCs and BGCs provides a very powerful improvement on the halo parameter uncertainties. The converging solution of the RGCs and BGCs considerably reduces (up to a factor of two) the 
area of parameter space enclosing the best-fit solution to the kinematics of the whole galaxy, including both the red and blue GCs. This makes GCs a most promising tool for testing \LCDM\ predictions with elliptical galaxy dynamics. The use of GCs has allowed us to confidently resolve most of the degeneracies involved and draw robust conclusions about the dark halo properties, the IMF of the host galaxy, and the anisotropy distribution of the two GC subpopulations.

In the future, we plan to extend this analysis to more systems, and possibly to lower masses, 
in order to investigate whether or not the high precision in halo parameters derived using the RGC/BGC 
joint analysis can distinguish Cold from Warm dark matter.   

\section*{Acknowledgments}

We thank the anonymous referee for useful comments which allowed us to improve the paper. DAF thanks the ARC for financial support from the grant DP130100388. This work was supported by the National Science Foundation through grants AST-0909237 and AST-1211995.  We thank Alis Deason for providing the mass profile of the galaxy in tabulated form.

\appendix

\section[]{Testing the Kurtosis accuracy using Monte Carlo simulations}\label{app:montecarlo}
\begin{figure}
\centering
\psfig{file=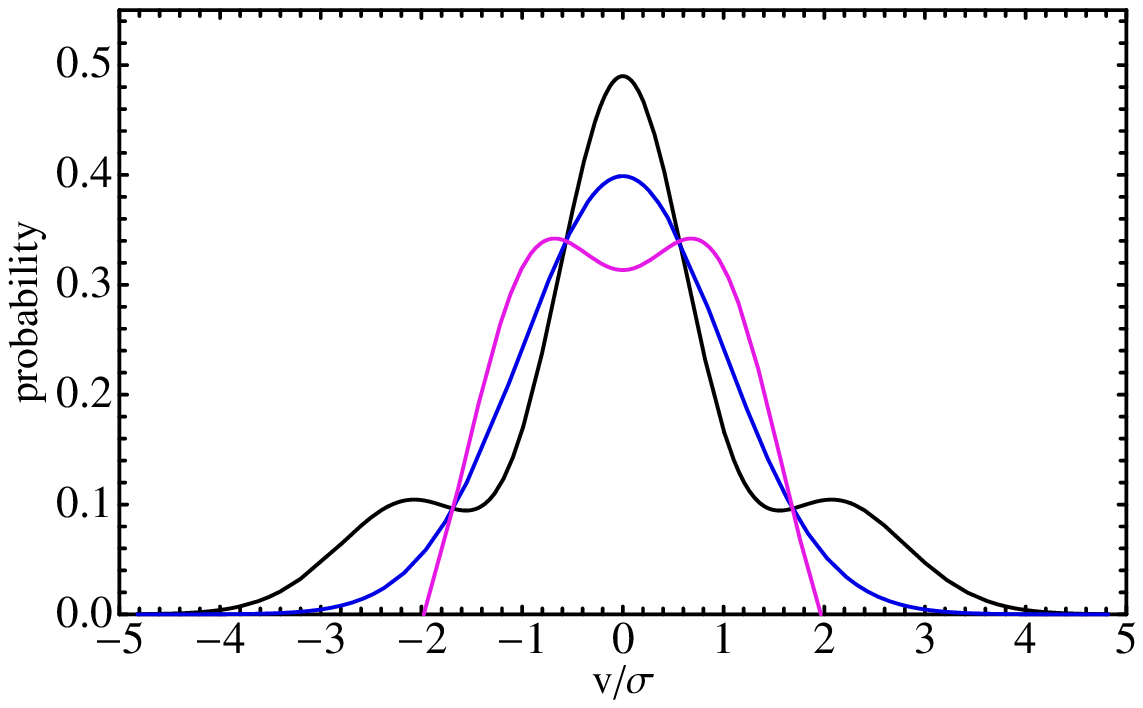,width=0.49\textwidth}
\psfig{file=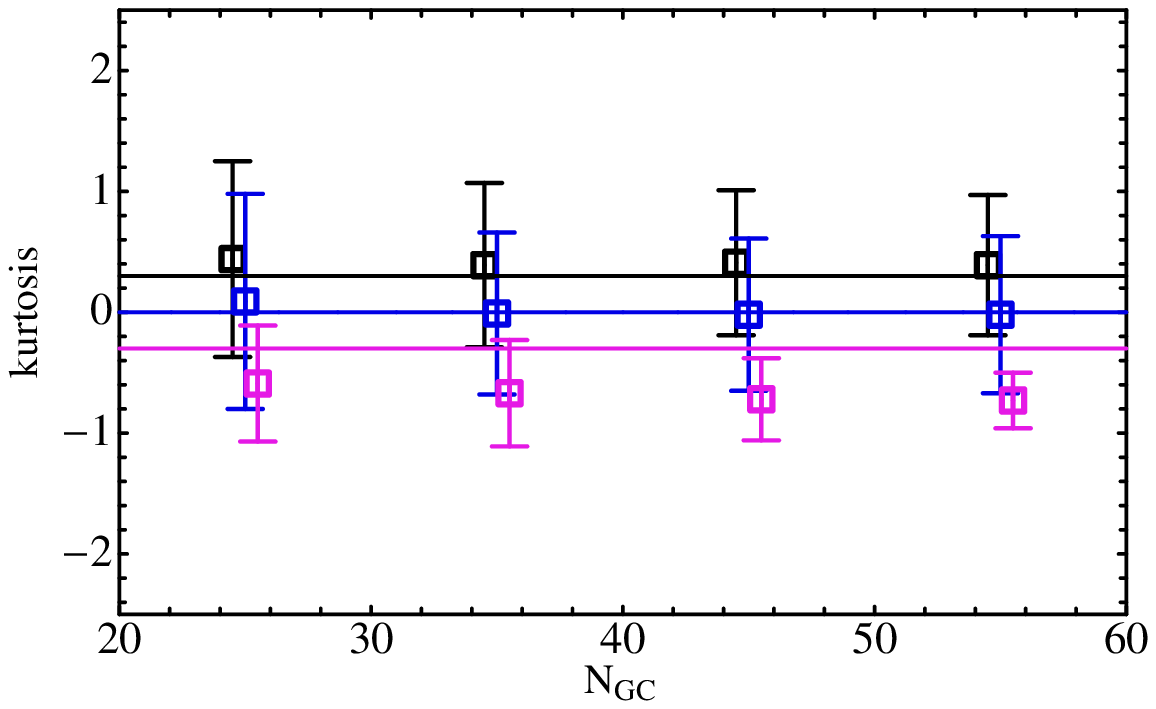,width=0.49\textwidth}
\caption{
Monte Carlo simulations of kurtosis measurements for different normalised 
line-of-sight velocity distributions (upper panel) corresponding to $\kappa=-0.3,~0,~0.3$ 
(purple, blue and black, respectively). The mean and standard deviation of the kurtosis 
estimates for different sample sizes and $\kappa$ over 200 random iterations are shown with 
same colours (lower panel). A small horizontal offset is adopted around the nominal $N_{\rm GC}$ 
to improve the readability of results.}
\label{fig:sim}
\end{figure}

As shown in \S\ref{sec:kinem}, the stringent colour selection adopted to ensure the cleanest
sample separation produces low number statistics in each spatial
bin. Bin sizes may be too small to guarantee robust results 
even when using an {\it unbiased} kurtosis definition (e.g. that of \citealt{J&G95}):
$$
\kappa=\frac{\overline{v_{\rm los}^4}}{\sigma_p^4}=\frac{n(n+1)}{(n-1)(n-2)(n-3)}\frac{\Sigma_i(v_i-\overline{v})^4}{\sigma_p^4}-3\frac{(n-1)^2}{(n-2)(n-3)}
$$
where we adopt the following unbiased definition for $\sigma_p$:
$$
\sigma_p^2=\frac{\Sigma_i(v_i-\overline{v})^2}{n-1}-\sigma_{\rm meas}^2,
$$
with $\sigma_{\rm meas}$ the individual velocity uncertainty (see P+13)
and $n$ the number of test particles.
The problem with unbiased estimators of dispersion and kurtosis
is addressed in \citet{2003MNRAS.343..401L}, where Monte Carlo simulations were used under the assumption of
a Gaussian parent distribution from which a given sample of objects is drawn
in order to define unbiased estimators.
Here we want to use a different approach by testing the robustness of the 
estimators drawn from non-Gaussian distributions.
Following \citet{1993ApJ...407..525V}, we assume symmetric (i.e. zero skewness) line-of-sight velocity distribution 
of the form
\begin{equation}
L(v)\propto\frac{e^{-v^2/2\sigma^2}}{\sqrt{2\pi}\sigma}[1+h_4 H_4(v)],
\end{equation}
where $H_4$ is the fourth Gauss-Hermite polynomial and $h_4$ is the corresponding 
coefficient. This coefficient is related to the classical kurtosis $\kappa\simeq 3+ 8\sqrt{6} h_4$
(or simply $8\sqrt{6} h_4$ for the excess kurtosis adopted here, see 
\S\ref{sec:kinem}). 
In Fig. \ref{fig:sim} (top) we show $L(v)$ for $\sigma=1$ and three different
$\kappa=-0.3,~0,~0.3$ spanning the typical kurtosis values reported in Figs. \ref{fig:fig_kin} and 
\ref{fig:fig_mods_2}. 
 
To evaluate the effect of the sample size on the kurtosis estimates of the
(intrinsic) line-of-sight distribution as above, we have 
randomly drawn $N_{\rm GC}=25,~35,~45,~55$ radial velocities from the above $L(v)$
over 200 Monte Carlo realisations 
of the same sample size. For each realisation, we compute the mean and standard deviation of the 
kurtosis estimated using the unbiased definition as above.   

The results are reported in Fig. \ref{fig:sim} (bottom) for the three $L(v)$. 
The unbiased definition seems to work fairly well
even for small samples at the cost of larger statistical uncertainties. 
However, the uncertainties are consistent with the observed uncertainties as defined in \S\ref{sec:kinem}. 
For negative kurtosis values, there is a significant bias toward more negative
values, which is even more severe for larger samples. Taking
these discrepancies at face value, the negative kurtosis for the BGCs might underestimate the true value that would
be more consistent with the model predictions of Fig. \ref{fig:fig_mods_2} (right panel).

The large statistical scatter of the kurtosis estimates might be a severe problem 
for our analysis relying on individual kurtosis profiles as derived in \S\ref{sec:kinem}. 
Thus, we attempt to assess the robustness of the profile from the GC data and the related 
error budget by performing a {\it Jackknife} resampling test, which consists of randomly 
drawing 3, 4 and 5 objects per 20--25 GCs bin in turn and recomputing the 
kurtosis before comparing the mean, 10th and 90th percentiles with the original 
estimates over 100 realisations. 
We find this {\it Jackknife} mean to be stable around the original values within 0.03. 
The 10th/90th percentiles are always smaller by $\sim70\%$ than the 
standard uncertainties of the kurtosis. Based on these results, we conservatively choose 
to use the larger error bars.

\bibliographystyle{mn2e}
\bibliography{napolitano_R2.bbl}

\end{document}